# BotNet Detection on Social Media

1st Aniket Chandrakant Devle
adevle@asu.edu

2nd Julia Ann Jose
jajose2@asu.edu

3rd Abhay Shrinivas Saraswathula
asarasw2@asu.edu

4th Shubham Mehta
smehta24@asu.edu

5th Siddhant Srivastava
ssriva59@asu.edu

6th Sirisha Kona
skona@asu.edu

7th Sudheera Daggumalli
sdagguma@asu.edu

**ABSTRACT**

As our reliance on social media platforms and web services increase day by day, exploiters view these platforms as an opportunity to manipulate our thoughts and actions. These platforms have become an open playground for social bot accounts. Social bots not only learn human conversations, manners, and presence but also manipulate public opinion, act as scammers, manipulate stock markets, and so on. There has been evidence of bots manipulating people's opinions and thoughts which can be a great threat to democracy. Identification and prevention of such campaigns that release or create these bots have become critical. Our goal in this paper is to leverage web mining techniques to help detect fake bots on social media platforms such as Twitter, thereby mitigating the spread of fake news.

Index Terms: Web Mining, Social Networks, Web Services, Disinformation, BotNets, Bot Detection, Coordinated Link Shar- ing Behaviour (CSLB), CoorNet, Community Detection.

## I. INTRODUCTION

Given the popularity of social media and the notion of it being a platform encouraging free speech, it has become an open playground for user (bot) accounts trying to manipulate other users using these platforms. Social bots not only learn human conversations, manners, and presence but also manipulate public opinion, act as scammers, manipulate stock markets, etc. Studies as in [1] have shown that platforms such as Facebook and Twitter are most affected by this phenomenon. Network of bots acting in this manner can pose a significant threat to the flow of data on social media platforms. Bot Networks (BotNets) can be involved in spreading malicious information, amplifying accounts or social domains on these platforms, thereby, exacerbating the issue of misinformation on social media platforms. This becomes especially dangerous in times of ongoing political crises. For example, Russian BotNets [2] were said to have played a huge role in the intense polarization between the Left and the Right during the 2020 US elections (until Microsoft stepped in). By doing so, bots play a huge role in the manipulation of thoughts of the citizens which is a threat to democracy. Our goal in this paper is to leverage semantic web mining techniques to detect fake bots and bot networks on Twitter.

## II. PROBLEM STATEMENT

The major objective of this paper is to develop a system that will detect fake accounts on social media platforms so that we get an honest and authentic feed. As the number of bots on the internet keep increasing day by day, it becomes even more cumbersome to detect and flag accounts part of a bot network. Bots these days are created using sophisticated machine learning algorithms and are made to sound very much like their human counterparts. Our goal is to detect bots accounts and BotNets and, further, detect the twitter accounts and domains that are amplified by these bot accounts and bot networks.

## III. RELATED WORKS

Many algorithms have been proposed for Fake/BotNet Detection on Social Media Platforms. Below are summaries of some of the existing algorithms:

1) A major differentiating attribute of online problematic information and malicious users is coordination. According to [3], Coordination is defined as the act of making people and/or things be involved in an organized activity together. In the papers [4]–[6] Coordinated Link Sharing Behaviour (CSLB) is used to detect entities that were involved in exhibiting inauthentic behavior. CLSB refers to the coordinated actions of Facebook/Twitter accounts, groups, and verified public profiles that posted the same news articles within a short period, showing a simple strategy for increasing content reach and gaming the algorithm that governs the dissemination of the most common content across the platform. In [5], authors considered political news stories of the 2018 Italian general election and 2019 European elections data. The detection of networks of coordinated entities is a two-step process. The first step is to estimate a time threshold to identify news items shared by different entities in a short period. The networks are then identified by grouping just the entities that repeatedly shared the same news story simultaneously.

2) In the paper *Identifying fake accounts on social networks based on graph analysis and classification algorithms* [7], Similarity matrices between accounts are calculated using graph adjacency matrix and then PCA algorithm is used for feature extraction and SMOTE is used for data balancing. To classify the nodes, linear SVM, Medium Gaussian SVM, regression, and logistic algorithms are used. The performance of this method was evaluated using the above classifier algorithms by training them on 10-fold cross-validation. Medium Gaussian SVM outperformed linear SVM due to its ability to map data to higher-dimensional feature spaces.

3) In Attractor+ [8], the paper focuses on exploring synchronized and coordinated retweeting behavior of malicious retweeter groups in terms of temporal and content-based properties. They proposed and built detectors based on the group-based features. They used four subgraph detection algorithms (Cohesive, Louvain, Attractor, Attractor+) to extract retweeter groups. They found that malicious retweeter groups had short and similar inter retweeting times (IRT) inspired by [9] by plotting IRT pairs on a log-log scale.
4) Enhanced PeerHunter [10] describes Peer-to-Peer BotNet detection using Community Detection strategies by clustering bots with "mutual contacts" into communities and finally, using a network-flow level community behavior analysis to detect BotNets. The main intuition here is that bots within the same BotNet will work together as a community and have features which can be distinguished from other communities.

IV. SYSTEM ARCHITECTURE AND ALGORITHM

The Semantic Web is an extension of the World Wide Web through standards set by the World Wide Web Consortium (W3C). The goal of the Semantic Web is to make the internet data machine-readable [11] Web Mining is the application of data mining techniques to discover patterns from the World Wide Web [12] Semantic Web Mining aims at combining these two research areas, that is, Semantic Web and Web Mining [13]. Among the various forms of malware, BotNets are emerging as the most serious threat against cyber-security [14]. To counter this, we are using Semantic Web Mining Techniques.

The first step in the project is the dataset analysis which in turn helps with pre-processing. Next, the entire pre-processed dataset is passed to the CoorNet algorithm for the detection of CLSBs. The CoorNet algorithm [6] works in two phases. First, a subset containing the top 10% of tweet groups with the shortest time span between first and second retweets is identified. This phase yields a threshold of time in seconds (Threshold 1). If a tweet's time difference of first and second retweet falls under this threshold, then the entire tweet group (all the retweets of this tweet) can be said to be suspects of exhibiting coordinated behaviour.

In the second phase each pair of retweets in the subset, rendered during the first phase, is observed by building a bipartite graph. The structure of the bipartite graph is as follows, retweet IDs on the left bipartite and twitter account names on the right bipartite. An edge between the left and right bipartite exists if a twitter account name has retweeted a tweet with a given retweet ID. This helps in identifying the second threshold by calculating the median number of retweets used by 10% of quickest retweets to reach 50% of their total number of retweets (Threshold 2). The second threshold is the number of times a pair of accounts are exhibiting retweeting behaviour abnormally.

The algorithm also furnishes a highly coordinated entities graph to help with the visualization purposes. The highly coordinated graph inherently has the Louvain community detection model applied on it which helps visualize the different communities amplifying a certain tweet or the twitter account responsible for that tweet. Open source software Gephi is utilized to visualize the graphs furnished by CoorNet.

The next step is to identify the twitter accounts which have been most amplified by the detected BotNets. The idea here is that: for every tweet group which has been classified with coordinated behaviour, the number of retweets made on this tweet is counted. Finally, this is aggregated for every twitter account name and 10 twitter accounts with highest coordinated retweet activity are selected. This gives a list of twitter accounts most amplified by BotNets. Bot names for the second dataset are derived by applying the algorithm which was used for DS-1.

As part of a future scope, we want to implement Topic Modeling algorithms like Non-negative Matrix Factorization (NMF) to discover the abstract topics which occur in the collection of tweets shared by the BotNets. Further, using Sentiment Analysis techniques we can extract and discover the positive or negative intent of the BotNets.

V. DATASETS

We intend to leverage the following datasets for experimentation purposes:

1) **UK 2016 Election Dataset (DS-1)**: This is a comma-separated dataset (CSV) wherein there is information on 5,000k tweets that are shared or retweeted by entities (public or private accounts) on Twitter along with the exact timestamp. The dataset contains tweets related to two events of major political crises, the UK elections of 2016 on Brexit and the resignation of Boris Johnson in 2018. There are 2449940 tweets from 2016 and 2755714 tweets from 2018. The format of the dataset is such that each tuple describes a directed edge from entity A to entity B, B sharing or retweeting a tweet made by A. We use this dataset as the basis for our model, any other dataset with differing schema is first preprocessed to match the schema of this dataset.
2) **US Covid Vaccine Dataset (DS-2)**: This is a comma-separated dataset (CSV) which contains Twitter API responses of tweets and retweets made by entities (public and private accounts) on Twitter. To make this dataset compatible with the model, tweet ID, retweet ID, screen names of accounts who tweeted and retweeted a particular tweet, and timestamps of corresponding retweets have been extracted in the pre-processing step.

We also extracted the retweet text and URLs to model the detection of amplified domains. This dataset contains tweets 1132525 related to the US government's strategy to track the Covid vaccination status of immigrants. We have taken the tweets from Dec 5 2020 - Jan 27 2020 time period.

## VI. EVALUATIONS

We use the given dataset as input to the CoorNet algorithm which returns a highly connected graph representing coordinated link shares between users and we feed this fastest retweet graph to the community detection algorithms. We plan on evaluating the results thus obtained using the dataset containing the bot names as ground truth and aim to quantify the performance using parameters like Accuracy, F1-score, precision, recall, and confusion matrix.

### A. Metrics

Confusion matrix: It reports the number of True positives, False positives, true negatives, and false negatives [15] and is shown in 1.

|  | Actual Class Positive | Actual Class Negative |
|---|---|---|
| Predicted Class Positive | True Positive | False Positive |
| Predicted Class Negative | False Negative | True Negative |

Fig. 1. Confusion matrix

Precision: It can be defined as the fraction of retrieved bots that are actually bot names [15]

Recall: It can be defined as the fraction of results that were successfully retrieved [15]

F1-score: It is merely a function of precision and recall that is used to seek a balance between the two.

### B. Experiments

For our experiments, we chose to divide the 2016 UK Election dataset into two sets of tweets, the first set consisting of all tweets related to the UK elections in 2016 (DS-1 2016 dataset), the second set consisting of all tweets related to the resignation of Boris Johnson (DS-1 2018 dataset). In short, we divided the dataset into two experiments and aimed to detect BotNets on these two sets independently.
Bots are predicted based on two approaches:
1) Tier-1: Accounts which satisfy the Threshold 1 and Threshold 2.
2) Tier-2: There could be cases where the accounts might not retweet within the shortest time span interval but perform coordinated retweeting behaviour.

To classify both Tier-1 and Tier-2 bots, we performed experiments by varying the second threshold.

DS-1 2016 dataset: Threshold-1 was 13 seconds, and metrics calculated are shown in Figure 2

| Threshold-2 (#retweets) | Total bots predicted | Bots predicted correctly | Total Bots in 2016 data | Accuracy | Precision | Recall | F1-score |
|---|---|---|---|---|---|---|---|
| 15 | 4244 | 158 | 313 | 98.78% | 3.72% | 50.4% | 6.92% |
| 30 | 2246 | 110 | 313 | 98.06% | 4.89% | 35.14% | 8.41% |
| 12 | 5460 | 181 | 313 | 98.58% | 3.31% | 57.82% | 6.24% |
| 10 | 6698 | 193 | 313 | 98.12% | 2.88% | 61.66% | 5.50% |

Fig. 2. Experiments and corresponding metrics for DS-1 2016 (UK 2016 Election Dataset)

DS-1 2018 dataset: Threshold-1 was 18 seconds, and metrics calculated are shown in Figure 3

| Threshold-2 (#retweets) | Total bots predicted | Bots predicted correctly | Total Bots in 2016 data | Accuracy | Precision | Recall | F1-score |
|---|---|---|---|---|---|---|---|
| 10 | 14674 | 407 | 546 | 94.03% | 3.03% | 74.54% | 5.84% |
| 15 | 10269 | 378 | 546 | 95.80% | 3.65% | 69.23% | 6.93% |
| 20 | 7906 | 336 | 546 | 96.77% | 5.05% | 61.53% | 9.35% |

Fig. 3. Experiments and corresponding metrics for DS-1 2018 (UK 2016 Election Dataset)

From the experiments and results in Figure 2 and Figure 3, we finalized the values as seen in the Figure 4.

| Dataset name | Threshold 1 (in sec) | Threshold 2 (frequency) | Total Bots in data | Total bots predicted | Bots predicted correctly | Accuracy | Precision | Recall | F1 |
|---|---|---|---|---|---|---|---|---|---|
| DS-1 2016 | 13 | 12 | 313 | 5460 | 181 | 98.58% | 3.31% | 57.82% | 6.24% |
| DS-1 2018 | 18 | 10 | 546 | 14674 | 407 | 94.03% | 3.03% | 74.54% | 5.84% |

Fig. 4. Accuracy, Precision, Recall, and F1 scores of the model on DS-1 (UK 2016 Election Dataset)

Figure 5 shows the metrics obtained on the 2nd Dataset (US Covid Vaccine Dataset).

| Dataset name | Threshold 1 (in sec) | Threshold 2 (frequency) | Total bots predicted |
|---|---|---|---|
| DS-2 | 55 | 10 | 4388 |

Fig. 5. Results metrics of the model on DS-2 (US Covid Vaccine Dataset)

### C. Findings

1) We realized quite early that the 2016 UK Election dataset was quite biased, the number of bots in comparison to the number of humans is quite small. Hence, a number of human accounts were classified as suspicious of coordinated behaviour during the first phase of CoorNet. But this was expected, in the sense that, so far we were only classifying bots based on the amount of time it takes for an account to retweet.

2) The second phase of CoorNet was able to weed out the majority of the bots from human accounts.

3) CoorNet also builds a highly connected graph of the input dataset and, using Gephi, we were able to visualize the graph. It is worth mentioning that most of the BotNets form a closed community (strongly connected component) of their own and work in tandem to amplify one or more twitter accounts.

4) Even though CoorNet performed relatively well on our dataset, we noticed some shortcomings on the overall model performance. First, CoorNet specifically targets fastest repetitive retweeters on a dataset. The handler of a BotNet can easily outwit our model by relaxing the time interval after which the bots have to retweet, hence, masking the overall BotNet from our model. Secondly, CoorNet specifically targets BotNets which form strongly connected components or disjoint communities in a network. Again, the handler can configure his bots to be part of different communities and still mask the overall BotNet from our model.

5) For dataset 1, the top 4 most amplified accounts were: Nigel Farage, Vote leave, leave EU and carolecadwalladr.

   a) Nigel Farage: British political commentator, broadcaster and former politician who served as Leader of the UK Independence Party. Farage was a key figurehead in the Brexit campaign of 2016/18, which, with 52 per cent of the vote, won
   b) Vote leave: Vote Leave is a campaigning organisation that supported a "Leave" vote in the 2016 United Kingdom European Union membership referendum.
   c) Leave EU: Leave EU is a political campaign group that was first established to support the United Kingdom's withdrawal from the European Union in the June 2016 referendum.
   d) carolecadwalladr: British author, investigative journalist and features writer. Wrote a series of articles for The Observer on the "right-wing fake news ecosystem."

6) It is worth noting that even though the accuracy of our model is very high, it is not the right metric to describe accuracy for this particular use case. The reason for such a high accuracy is mainly due to the huge number of human accounts that CoorNet has classified correctly as compared to the number of bot accounts. Hence, recall is a better metric to describe the accuracy of CoorNet.

## VII. UI/Visualization Interface Designs

We used Gephi to visualize the highly connected graphs generated by CoorNet for analyses. Also, we have used Python's standard libraries like MatPlotLib to generate pie charts depicting the amount of amplification on the top 10 amplified twitter accounts by bots.

In figures 6, 7, and 8, the vertices correspond to various twitter accounts. Also, the nodes are sized on the basis of their degree, the larger the degree of the node, the larger it appears. It is also notable that larger degree means a particular twitter account is amplified by a larger number of accounts, bots and humans alike. However BotNets appear to form highly connected communities as evident in the graphs.

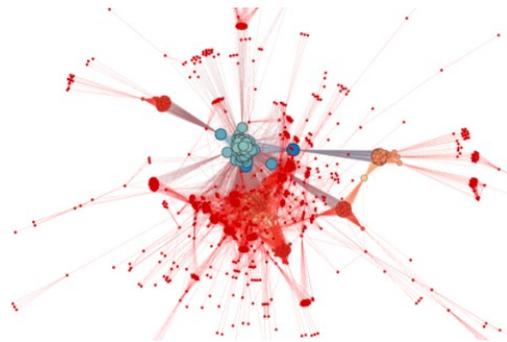

Fig. 6. Graphical visualization of DS-1 2016 (UK 2016 Election Dataset)

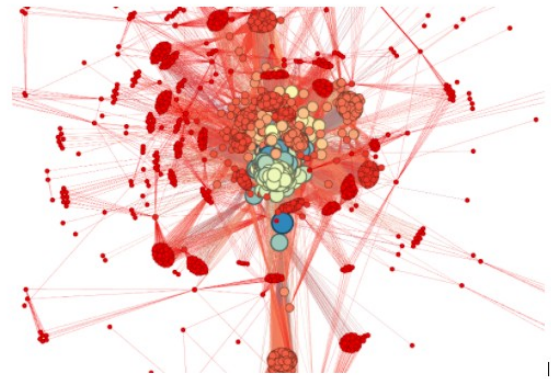

Fig. 7. Graphical visualization of DS-1 2018 (UK 2016 Election Dataset)

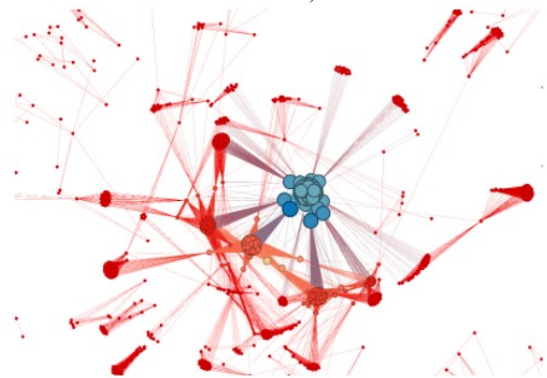

Fig. 8. Graphical visualization of DS-2 (US Covid Vaccine Dataset)

Figures 9, and 10, mentions the 10 most amplified twitter accounts by BotNets along with their respective shares of amplification for DS-1 2016 and 2018 datasets. It is worth mentioning that these accounts were major players during the political polarization in the UK in 2016 and 2018, each being at the forefront of for or against Brexit. Figure 11 mentions

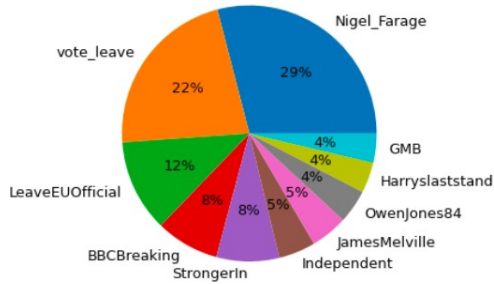

Fig. 9. Pie Chart showing most boosted accounts for DS-1 2016 (UK 2016 Election Dataset)

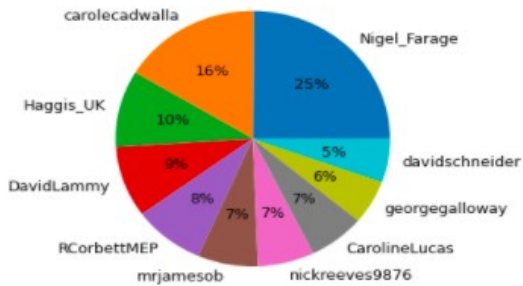

Fig. 10. Pie Chart showing most boosted accounts for DS-1 2018 (UK 2016 Election Dataset)

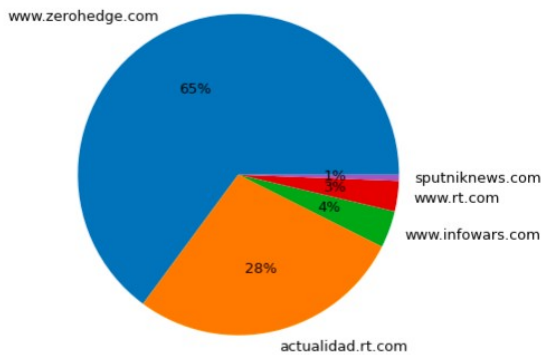

Fig. 11. Pie Chart showing most boosted domains for DS-2 (US Covid Vaccine Dataset)

the most boosted domains for DS-2 and Figure 12 mentions the most boosted accounts for DS-2.

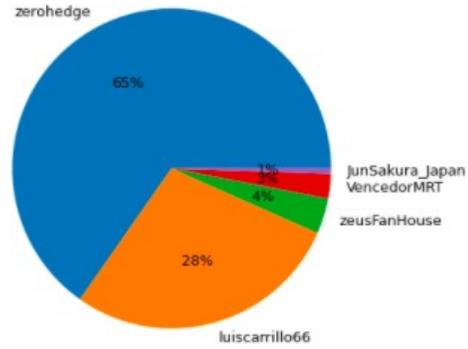

Fig. 12. Pie Chart showing most boosted accounts for DS-2 (US Covid Vaccine Dataset)
.

## IX. CONCLUSIONS

In this paper, we covered the CoorNet BotNet detection model, which works on the principles of Coordinated Link Sharing Behaviour (CLSB). CLSB suggests that it is unlikely for a human to share content on any social media platform repeatedly within a small time threshold. We also covered various different BotNet detection algorithms like Enhanced PeerHunter [10] and Attractor+ [8] with varying complexities and results.

It should be pointed out that bots keep getting better and better and it is still a challenge to develop models and algorithms that effectively detect BotNets in all kinds of configurations. We mentioned some of the shortcomings of CoorNet as well, that it will falter if BotNets become a bit more sentient about their coordinated behaviour.

While implementing CoorNet, we applied thresholds to detect coordinated behaviour in two tiers. In the first tier, we looked at the top 10% fastest retweeters in the first phase and then captured the accounts with highest retweeters in the second phase. In the second tier, we directly applied the idea of phase two from tier one to capture bots who were missed out by tier one, owing to the lower number of retweeters.

We generated highly connected graphs for each of our datasets to observe and analyse the configuration of BotNets in a network and realised that BotNets effectively form strong disjoint communities. We also analysed the top accounts which were amplified by BotNets for each of the datasets and the top domains which were boosted by BotNets for DS-2. By experimenting on the second threshold for tiers one and two, we were able to achieve a recall of 57.82% for DS-1 2016 dataset and 74.54% for DS-2 2018 dataset.

The model can still be further improved and as part of future scope, we want to add topic modelling and sentiment analysis features into the model to detect the abstract topics amplified by BotNets and their positive or negative intentions.

# REFERENCES

[1] P. Podvin, The Social Impact of Bad Bots and What to do About Them. https://www.forbes.com/sites/forbestechcouncil/2020/12/04/the-social-impact-of-bad-bots-and-what-to-do-about-them/?sh=2265503259e0 (accessed 7th February 2021)

[2] J. Greene, Microsoft seeks to disrupt Russian BotNet it Fears Could Seek to Sow Confusion in the Presidential Elections, The Washington Post. https://www.washingtonpost.com/technology/2020/10/12/microsoft-trickbot-ransomware/ (accessed 7th February 2021)

[3] N. Maréchal, "Gabriella Coleman, Hacker, Hoaxer, Whistleblower, Spy: The Many Faces of Anonymous." International Journal of Communica- tion [Online], 2015.

[4] F. Giglietto, N. Righetti, L. Rossi, and G. Marino, " Coordinated link sharing behavior as a signal to surface sources of problematic infor- mation on facebook", International Conference on Social Media and Society, pp. 85-91, July 2020

[5] F. Giglietto , N. Righetti , L. Rossi G. Marino, "It takes a village to manipulate the media: coordinated link sharing behavior during 2018 and 2019 Italian elections", Information, Communication, and Society, 23:6, pp. 867-891, 2020.

[6] F. Giglietto, N. Righettiand G. Marino, "Understanding Coordinated and Inauthentic Link Sharing Behavior on Facebook in the Run-up to 2018 General Election and 2019 European Election in Italy", 20-Sep-2019. [Online]. Available: osf.io/preprints/socarxiv/3jteh.

[7] M. Mohammadrezaei, M. Shiri and A. Rahmani. "Identifying Fake Ac- counts on Social Networks Based on Graph Analysis and Classification Algorithms." Secur. Commun. Networks 2018, 2018.

[8] No. Vo, K. Lee, C. Cao, T. Tran, and H. Choi, "Revealing and detecting malicious retweeter groups", IEEE/ACM International Conference on Advances in Social Networks Analysis and Mining (ASONAM), pp. 363-368, IEEE, July 2017.

[9] A. Costa , Y. Yamaguchi, A. Juci Machado Traina, C. Traina Jr, and C. Faloutsos, "Mining and modeling temporal activity in social media" Proceedings of the 21th ACM SIGKDD international conference on knowledge discovery and data mining, pp. 269-278, August 2015.

[10] D. Zhuang, and J. M. Chang, "Enhanced peerhunter: Detecting peer-to-peer botnets through network-flow level community behavior analysis.", IEEE Transactions on Information Forensics and Security, 14(6), 1485- 1500, 2018.

[11] Semantic Web, Scientific American. https://www.scientificamerican.com/article/the-semantic-web/ (accessed 7th February 2021)

[12] "Web Mining", GeeksforGeeks, https://www.geeksforgeeks.org/web-mining/ (accessed on 10th February 2021)

[13] B. Berendt, A. Hotho, G. Stumme, "Towards Semantic Web Mining", The Semantic Web ISWC 2002, Springer, Berlin, Heidelberg, 2002, vol 2342.

[14] M. Feily, A. Shahrestani and S. Ramadass, "A Survey of Botnet and Botnet Detection," 2009 Third International Conference on Emerging Security Information, Systems and Technologies, Athens, Greece, 2009, pp. 268-273, doi: 10.1109/SECURWARE.2009.48.

[15] K. P. Shung, "Accuracy, Precision, Recall, or F1?", towardsdata-science. com, https://towardsdatascience.com/accuracy-precision-recall-orf1-331fb37c5cb9 (accessed on 21st February 2021)